# The River and the Sky: Astronomy and Topography in Caral Society, America's First Urban Centers

A. César González-García, Aldemar Crispín, Ruth Shady Solís, José Ricra, Felipe Criado-Boado, and Juan A. Belmonte

**PREPRINT**

**A. César González-García** and **Felipe Criado-Boado** Instituto de Ciencias del Patrimonio (Incipit), Consejo Superior de Investigaciones Científicas (CSIC), Avda. de Vigo s/n, 15705, Santiago de Compostela, Spain (a.cesar.gonzalez-garcia@incipit.csic.es, corresponding author)

**Aldemar Crispín** Zona Arqueológica Caral, Ministerio de Cultura, Lima, Perú

**Ruth Shady Solís** Zona Arqueológica Caral, Ministerio de Cultura, Lima, Perú, and Facultad de Ciencias Sociales, Universidad Nacional Mayor de San Marcos, Lima, Perú

**José Ricra** Instituto Geofísico del Perú, Lima, Perú and Grupo de Astronomía, Facultad de Ciencias, Universidad Nacional de Ingeniería, Lima, Perú

**Juan A. Belmonte** Instituto de Astrofísica de Canarias, Calle Vía Láctea s/n, 38200, La Laguna, Tenerife, Spain and Departamento de Astrofísica, Universidad de La Laguna, Spain

America's first urban centers were allegedly located at the Supe Valley sites in Peru. After investigating the location and the orientation of the main built structures, we show that it is not only the presence of the River Supe that determines their orientation but also astronomical relationships within the orientation of the buildings dictate their setting within the valley. The southernmost position of moonrise on the horizon seems to be the most important astronomical target. There is the possibility of an evolution toward attributing greater importance to the June solstice sunrise and the rising of certain stars or asterisms. These orientations could relate to specific moments throughout the year, in particular to seasonal rains, subsequent river flooding, and agricultural cycles. This is one of the earliest examples of the interaction of land- and skyscapes in human cultures and indeed the first in the Americas.

El valle del río Supe en Perú posiblemente alberga una de las primeras manifestaciones de urbanismo en América. En este artículo investigamos la localización y orientación de los edificios principales de esta cultura. Los resultados muestran que la presencia del río Supe determina de forma clara la orientación de estos edificios, pero la localización dentro del valle también viene dictada por posibles relaciones astronómicas. En concreto, se muestra que las orientaciones de estas estructuras concuerdan de forma fehaciente con la salida más meridional de la luna. Existe a su vez la posibilidad de una evolución que incluya una importancia creciente del solsticio de junio y la salida de ciertas estrellas y asterismos. Estas orientaciones se pueden relacionar con momentos concretos a lo largo del ciclo anual, en particular con las estaciones húmedas, las crecidas del río y los ciclos agrícolas. De esta manera, los monumentos del valle del Supe aparecen como una de las primeras muestras de interacción del paisaje y el celaje en América.

The settlements of the Initial Formative and Early Formative periods (3000–1500 BC; Shady et al. 2001, 2015) of Caral society are located in the Supe and Huaura Valleys in the north-central area of Peru. These sites, contemporary with others in nearby valleys of the central coast of Peru, experienced substantial cultural change, including the emergence of arguably the first complex societies in the Americas. The settlements of the Supe Valley show greater complexity and marked differences in extension and volume from others in the area (Burger and Rosenswig 2012; Shady 2014a). They have some of the earliest examples of monumentality in the Americas and were the first to be petrified, using intensive stone works arranged in complex urban-like centers (Haas and Creamer 2012) that conformed to what has been proposed as the first urban civilization in the Americas (Shady 2006a, 2014b). These early architectonic developments, which include both public buildings and residential areas, required a high degree of logistical complexity that involved not only complex organizations but also, as is demonstrated here, a well-developed spatial and astronomical awareness.

At both coastal and inland settlements, animal protein (90% of all protein according to Burger and Rosenswig 2012; see also Béarez and Miranda 2003; Shady 2006a, 2014a) came from the predominant maritime species, the anchoveta (*Engraulis ringens*). The farmers used irrigation in the valley to grow cotton (*Gossypium barbadense*), which was used for making clothes and fishing nets (Béarez and Miranda 2003; Burger and Rosenswig 2012; Shady 2006a, 2014a). They also grew gourds (*Lagenaria siceraria*), which were also in great demand in preceramic societies for serving and storing food. Other species identified include pumpkin and *Cucurbita* (squash), beans (*Phaseolus vulgaris*), arrowroot (*Canna indica*), and sweet potatoes (*Ipomoea batatas*; Béarez and Miranda 2003; Burger and Rosenswig 2012; Sandweiss et al. 2009; Shady 2006b; Shady and Leyva 2003). Exchange or trade with other regions is evident from the frequent finds of items from elsewhere, such as *Spondylus* shells from the equatorial seas (Burger and Rosenswig 2012; Shady 2006a, 2014a). These finds, together with the material culture recovered, indicate a complex society that was able to mobilize and plan elaborate building projects involving an extensive labor force, with an economy based on fishing, irrigated agriculture, an urban setting, and goods exchange or trade (Burger and Rosenswig 2012; Shady 2014a). After nearly one thousand years of splendor, the society abruptly ended, possibly because of several factors driven by environmental changes (Sandweiss et al. 2009; Shady 2006a).

Some 25 sites with monumental architecture (Figure 1) have been identified in the first 50 km from the seashore, inland into the valley (Burger and Rosenswig 2012; Shady 2006a, 2014a; Shady, Dolorier, et al. 2000). These sites share the same settlement pattern. They appear on both sides of the river, with roughly the same number on each side and grouped into sections: 9 in the medium-high section, 10 in the medium-low section, and 6 in the low section of the valley and the seashore, respectively. The inhabited space seems to follow a certain organization, and all the settlements have, on different scales, subsets of residential units and public monumental buildings (Burger and Rosenswig 2012; Shady 2006a, 2014a).

<Insert Figure 1 about here>

In general, the valley is formed by three terraces, designated TF0, TF1, and TF2. TF0 corresponds to the present-day river flow, TF1 appears at a variable height between 1 and 4 m above the river flow, and TF2 is 15–25 m above this level. It is on TF2 where most archaeological remains appear (Carlotto Caillaux et al. 2011). Each summer the River Supe fills all its flow and floods the alluvial plains. With extraordinary rains, the river floods

TF1. Currently TF0 seems to be mostly parallel to TF2. We can speculate that even if small shifts may have happened since the building of Caral and the other ceremonial centers, these should have occurred either within TF0 or at most TF1, which seems roughly parallel— although, in general, wider than TF0—to TF2.

Because most of the settlements were in these arid areas about 25 m above the valley floor, the constructions did not prevent extension of fertile and irrigable lands. The most extensive archaeological sites are found in the largest of these empty terraces. Interestingly, the largest centers, in terms of both area and constructed volume, are on the left bank of the medium-low section of the valley, where less farmland is available than on the right bank (Burger and Rosenswig 2012; Shady 2014a).

In particular, the site of Caral is outstanding for its large area, the design of its constructed space, the volume of its buildings, and its state of preservation. First visited by Kosak and Engel and originally called Chupacigarro/Caral, extensive early investigations carried out in the area identified other nearby settlements, named Chupacigarro Grande, Chupacigarro Chico, Chupacigarro Centro, and Chupacigarro Oeste (Engel 1987; Kosak 1965; Shady 1997; for a history of early investigations, see, for example, Benfer 2012; Haas and Creamer 2004). To differentiate the several sites, they were renamed, respectively, Caral, Chupacigarro, Miraya, and Lurihuasi. In this article, we follow the most recent nomenclature for the sites to avoid misinterpretation.

The constructions in Caral were apparently planned as a group, from the Ancient period toward 3000 BC (Burger and Rosenswig 2012; Sandweiss et al. 2009; Shady 2006a, 2014a; Shady et al. 2001). During the more than one thousand years of occupation, the buildings were periodically remodeled and enlarged. Their structural stability was gradually improved with reinforcement of the deposits of the platforms using *shicras* or woven bags as containers or gabions, allowing the growth of the construction in a pyramidal form (Shady 2014a; Shady et al. 2009). Around 2000 BC, in the Late period, labor investment in architecture began to decrease gradually (Burger and Rosenswig 2012; Shady 2006a; Shady et al. 2001).

Some authors argue that there are few domestic areas in comparison with the monumentality of the public buildings. This would suggest their use as religious and ceremonial centers, where the population gathered at special moments without engaging in real urban relations (Makowski 2006). Yet, the urban character of these centers should not solely rely on the nucleated structure of domestic areas, because it is now evident that dispersed urban centers do occur, especially in tropical areas (e.g., see Smith 2014:308 and references therein); one example might be Angkor in Cambodia; Carter et al. 2018).

A characteristic architectonic design often connects a pyramidal building with a sunken circular plaza (Shady 2006a; Shady, Machacuay, and Arambury 2000). This is a recurrent feature in the settlements of the Supe Valley and other neighboring valleys (Figure 2) and has been related to religious ideology (Burger and Rosenswig 2012; Piscitelli 2018).

<Insert Figure 2 about here>

## From Space to Time

In this context, it is important to realize that the monumental buildings created by this society may, in their location and spatial configuration, embed relevant information that may shed light on how the builders apprehended space and time (e.g., see Tilley 1984:121–125). Tilley defines places as centers with significant meaning for persons, whereas space can be understood as composed of places that create landscapes (Tilley 1994). In this sense, he interprets the architectural space as the deliberate creation of space or as the space constructed by a society and its social relations. Where the social interaction takes place is important; therefore, place is not neutral but is loaded with value and power as defined by that society. Finally, such a concept of space interestingly applies to where monuments are placed in the landscape and how they relate to their environment, how they "observe" and are observed, where they are facing, and their orientation. Thus, a direction of a monument has a concrete social meaning, and we can argue that the orientations embedded in a monument include valuable social information (Tilley 1984:122).

Finally, when such orientations are related to the recurrent and cyclical movements of heavenly bodies, then some sense of temporality can also be assumed, and we can argue that such conceptualization of time is also loaded with social value. In this sense, both space and time form part of social practices (see Bender 1998; Ingold 1993; Massey 2006).

Early work by Milla Villena (1983:153) mentions Chupacigarro/Caral as a site with potential archaeoastronomical relations. Although preliminary results at the Caral site (Benfer 2012; Benfer et al. 2007; Marroquín Rivera 2010; Pinasco Carella 2004) appear to suggest such relations in the orientation of its buildings, several methodological issues cast doubts on their conclusions. Primarily, most of these measurements did not consider the horizon altitude that in several instances can render this kind of approach fruitless. Benfer (2012, referring to an unpublished talk by the author in 2006) considered the horizon altitude and suggested some possibilities similar to the later site of Buena Vista. Benfer and others (2007) indicate that the amphitheater has an azimuth of 114º36' and relates this to the Milky Way, the summer solstice, or the lunistice (Benfer 2012). However, such a horizon is so close as to make it difficult to assess whether any of these possibilities could have been important in other areas of the same site. According to these authors, the main plaza could be related to the summer solstice. Second, all these works focused on a single site, Caral itself, and any conclusions should thus be considered as preliminary.

The aim of our research was to investigate the location and orientation of a comprehensive sample of buildings at several sites in the Supe Valley to determine whether such measurements could provide any insight into the spatial and temporal concepts of Caral society. We wanted to verify whether the river was the main driver vector that determined the orientation and disposition of the structures at the different sites and whether there was also a possible astronomical intentionality in those orientations.

We investigated a comprehensive set of buildings for two reasons. First, only a statistical analysis such as the one proposed here might be able, given the absence of contemporary records, to provide support for or against the possible existence of intentionality. Second, if such intentionality appears and is related to a given astronomical event, we could speculate on the ritual intent determining such a pattern.

## Fieldwork Methodology

During our field campaign in October 2016, we investigated 10 sites: eight are directly related to the River Supe, and two, Áspero and Vichama, are nearer to the seashore. Vichama is in the valley of the Huaura River and shares characteristics with the sites at the Supe Valley, being contemporary to Piedra Parada in the Early Formative period (Shady et al. 2015). These 10 places include all the sites that extend beyond 15 ha (Burger and Rosenswig 2012; Haas and Creamer 2012) and are thus the larger ones; they also include a comprehensive number of already excavated structures. In total, we analyzed 55 structures, with a total of 181 data points, in those 10 sites.

In every site, we measured the directions of the principal axes of the larger pyramidal buildings whether or not they were associated with circular sunken plazas. In most cases, this meant four directions, but for several structures, a smaller number was measured owing to their state of preservation. We did not consider the stairways as the principal direction at this stage, because in some cases, the several phases present stairways on different sides of the same building; most probably this indicates that the erection of the stairway was possibly more related to an urban planning decision (such as closeness to a central plaza) than to the building's ritual orientation. When different construction phases of the building were identified with variations in the orientation, we presented more than four data points for a particular building. The circular plazas often included only the orientation toward the stairs leading into the plaza, but in several instances where this direction was not clear or when other features appeared to be interesting, other measurements were also included, such as the perpendicular to the stairs. Although the largest site in our sample is Caral, it is worth stressing that this site does not directly dictate the overall results.

We took measurements of the four sides of each building using two professional-grade magnetic compasses: a Suunto Tandem 360R and a Silva SurveyMaster 360. Each measurement was corrected for magnetic declination using two methods. First, we employed a global Earth model, available at https://www.ngdc.noaa.gov/geomag-web/. Second. we verified the readings, using selected data points obtained by the team at Caral using total stations and double-checked with direct observation of sunrise on particular dates. This provided a robust handle on our measurement uncertainties, so that despite the fact that the compasses readings have a ¼° error in azimuth and ½° error in angular altitude of the horizon measured in the line of sight, the astronomical declination error would be less than ¾°.[1] Although such uncertainties might be important in ascertaining the correct date corresponding to a given direction, with an error of ± 1 day, given the kind of statistical analysis that we present here, we believe that this should pose no major problem.

We then converted the data obtained—the azimuth corrected for magnetic declination and horizon altitude—to astronomical declination for each site, considering the local latitude of the site and correcting the horizon altitude from atmospheric refraction (Table 1). To account for this effect, we employed a semi-empirical formula (Schaefer 1993). In the following, dates are given in the Gregorian proleptic calendar.

<Insert Table 1 about here>

## Results

Figure 3a shows the orientation diagram for all our measurements. The circle indicates all directions of the compass, the small strokes outside the circle indicate the cardinal directions, and DS and JS stand for the December and June solstices, respectively. Note the treble shape of the diagram. This occurs because we measured all directions in the pyramidal buildings and sunken plazas, so most are represented up to four times within the diagram. There are clear concentrations next to the DS sunrise and JS sunset, with an apparent wide dispersion of the data.

<Insert Figure 3 about here>

Figure 3b shows the results when considering the orientation of the buildings with respect to the Supe River for those sites directly related to it within the valley. The orientation of the river is taken from that given by the riverbed (i.e., the orientation of the contour lines of TF0) closest to each archaeological site—the obvious choice to test the hypothesis that the river dictated the orientation of the buildings—whereas, for each structure, we considered the azimuth closest to it. The riverbed might have changed since the time of construction of the Supe settlements, and in given spots it might have led to different orientations within the alluvial plain. However, the general layout of the valley imposes an orientation that is followed by TF2, on which the structures are built; although individual changes may have occurred, we argue that the overall orientation was generally maintained.

In the data analysis presented in Figure 4 and in the following discussion, we used a kernel density estimator (for an explanation, see Weglarczyk 2018) to create a probability density function. To do so, we used both an Epanechnikov and a Gaussian kernel with a bandwidth of twice the declination uncertainty, thereby including the error in the density estimation. The histogram is built as the sum of each of those kernels and thus gives the concentration of declinations.

Figure 3b indicates that the riverbed direction or one very close to it is a preferred factor to orient the structures. In other words, the placements of the largest pyramidal buildings for all sites in the Supe Valley associated with the river seem to be directly dictated by the riverbed. Yet, this does not mean that all the buildings are facing the river; in several cases, the stairs leading to the top part of the building do not face the river. However, as indicated earlier, the placement of the stairs seems to be dictated by urban planning reasons, because in most of these cases such stairs lead to an open area, a plaza.

However, the buildings' orientations seem to be affected by other factors as well: the orientation diagram (Figure 3a) indicates the possible relevance of astronomical directions. Given that we have four directions, we wanted first to verify whether there is any indication of a preferred one.

We performed a test to ascertain which direction should be more significant from an astronomical point of view (González-García and Sprajc 2016). The test compared the prominence of the concentration of orientations in a given directions arising from considering the four directions of each building (Figure 4). The hypothesis is that if the astronomical orientation is important toward a particular point in the sky as seen rising or setting, that would take into consideration both the orientation and the altitude of the horizon in one particular direction, but not in the perpendicular ones or in the one opposite to it. For a statistical sample, this would yield a sharper peak in one direction while producing shallower and wider concentrations in the other relevant directions. According to this scheme, which has four concentrations of values corresponding to an orientation group, the sharpest and highest peak most likely indicates the targeted value and therefore suggests the direction in which the orientations of buildings of this group were functional. To show this we compared the prominence with respect to the error introduced by using different kernels for the histograms.

<Insert Figure 4 about here>

We calculated the mean error introduced by using these two different kernels through the root mean square of the differences between the two kernels at the most prominent maxima in declination to consider the effect of the local horizon. This value for the curvigrams shown in Figure 4 is rms = 0.002.

Comparing the amplitudes of the maxima in the four directions, the distribution of declinations on the eastern horizon exhibits more pronounced concentrations. In agreement with our assumption, this distribution suggests that the orientations were intended to be functional predominantly in the eastern direction. As a result, the relevant direction is toward the eastern horizon for each building. Therefore, in subsequent figures we only use the eastern data.

To verify whether the concentrations in declination are statistically significant, we compared our results with those of a uniform distribution in azimuth located at the latitude of the Supe Valley with 80,000 data points. We extracted from such an ideal distribution 100 random distributions with as many elements as we considered in the data sample and compared the resultant distribution with the measured one.

Figure 5 presents the comparison between the eastern orientation data (dark-gray shadowed curvigram, which we call the measured distribution, f(obs)), and the 100 random distributions taken from a pool of 80,000 data points homogeneously distributed in azimuth between 45º and 135º (black curves). The variations observed between the 100 random distributions help us calculate the variations, σ(unif), with respect to the uniform distribution (white solid curve, f(unif)).

<Insert Figure 5 about here>

We verified the statistical significance of our results in two ways. The first method uses a Kolmogorov-Smirnoff test for each of those 100 comparisons. For all the tests performed, we obtained probability values of less than 0.05, where the Kolmogorov-Smirnoff test would allow discarding the null hypothesis that the observed distribution and the random distribution could be drawn from the same parent population.

The second method calculates a mean standard deviation of the random samples from the uniform parent distribution (σ(unif)). The difference between our measured data and the uniform one can then be scaled in sigma levels. In this way, we were able to verify the significance of the concentrations in the curvigrams. Figures 6 and 8 provide the observed data scaled according to the standard deviation (σ(unif)), which shows whether the concentrations are prominent. The measured distribution (f(obs)) is compared with the concentrations expected if the orientations were uniformly distributed to any points in the horizon (f(unif)), and scaled with respect to the variation expected in this distribution (σ(unif)). Any peaks larger than a value of 3 are regarded as statistically significant. A first version of this method is introduced in González-García and Sprajc (2016:192–193).

Figure 6a shows the orientation diagram similar to that presented in Figure 3a, but only with those measurements toward the eastern sector of the horizon (i.e., between 45° and 135° in azimuth; this is justified by the exercise described earlier to determine the most relevant direction). It emphasizes the wide concentration toward the area of the DS. We have 73 measurements toward the eastern sector (Figure 6b), and the largest concentration in declination appears toward the major lunistice; that is, if we consider the full moon as the relevant lunar phase, this would be the full moon just before or after the JS.[2] A second maximum appears at declination −20º, close to the minor lunastice, and there seems to be a third significant peak close to the DS sunrise.

<Insert Figure 6 about here>

The sunken plazas and their associated buildings seem to present slightly different orientation patterns than other buildings. When the sunken plazas are considered separately (Figure 6c), the lunar peaks at −29º and −18º are emphasized. Alternatively, the second largest maxima, especially for the sunken plazas, could be related to the rise of Sirius for the epoch of use of the different areas, as shown in Figure 2d–f for Vichama building A1. Other examples are building H1 of Lurihuasi, C1 in Miraya, A from Chupacigarro, and A1 from Piedra Parada (see Table 1; the change in declination of Sirius is marked in Figure 6c for the last period of occupation of the settlements in the Supe Valley). In addition, a third maximum appears (Figure 6c) at declination −34 º; although it is not above the 3-sigma level, it might be related to the rise of the Southern Cross. Arguably, the western constellation was not recognized by ancient sky watchers; however, it was perhaps recognized as important in its association with the dark clouds of the Milky Way. Interestingly, much later in time this part of the sky was the *yakana* recognized by the Quechuas—the Llama constellation, which was a very relevant asterism in Andean sky folklore (Pucher de Kroll 1950; Taylor 1987; Urton 1981; Zuidema 1982). Finally, the orientation of the remaining buildings follows a similar pattern (Fig 6d), but there is no concentration at −18º and we could not discard the possibility that certain structures could also be related to the DS sunrise. This is also illustrated in Figure 7.

<Insert Figure 7 about here>

## Discussion

These results suggest that the course of the riverbed dictated the overall orientation of the built structures for the Supe culture. At the same time, the interaction between the local landscape and related astronomical phenomena played a role. The river does not follow the same general direction throughout the area in which the investigated sites are located, perhaps indicating that the sites were deliberately located on the spots where astronomically relevant directions were parallel or perpendicular to the riverbed. In some instances, there is the impression that several of those pyramidal buildings were mirror images of the mountains behind them, as seen from the open spaces (*plazas*) next to them. However, we did not verify this for all the measured buildings, because that task was beyond the scope of the present study.

This would therefore be the earliest example in the Americas of a close relationship between land- and skyscape in a way that echoes what happened in contemporaneous cultures of the Old World, such as ancient Egypt (Belmonte et al. 2009). A similar situation could be identified in Ecuador (Ziedler 1998) and roughly two millennia later in coastal Peru and elsewhere (Malville 2015a), with Chankillo as a paradigm (Ghezzi and Ruggles 2007, 2015)—reaching its most complex and prominent level in the Vilcanota/Urubamba "sacred" Valley in the Andes in Inca times (for a recent review, see Malville 2015b;

Ziolkowski 2015). The sun seemed to have played a dominant role in those areas. Thus, the river and the sky influenced the layout of public buildings (Adkins and Benfer 2009; Ziedler 1998). Other analogous examples could be found in Mesoamerica, where many important buildings were oriented both on astronomical grounds and with respect to prominent mountaintops on the local horizon; the orientations of most temples on the northeast coast of Yucatan peninsula, for example, conform to the adjacent shorelines but also pertain to astronomically significant groups (Sprajc 2018:218f, and references therein).

However, a peculiarity of the Supe Valley culture is the importance of the moon and its extreme positions as the apparently dominant elements in the orientation of sacred structures within the different settlements—notably but not only the circular sunken plazas—and eventually of very important stars and asterisms such as Sirius or the Southern Cross. The Southern Cross could perhaps be linked with the dark constellation of the Llama. Benfer and others (2007) also argue for the importance of such constellations in Buena Vista (for them it is the Fox constellation). However, we have no relevant cultural information that such a constellation was recognized in these early times. Therefore, any speculation must be made cautiously given the large span of time from the remains analyzed here and the evidence we have regarding the importance of such constellations for the region. Indeed, this is an open question for further research.

The relevance of the moon and lunar cults has also been claimed for Buena Vista at the Chillón River for the period after Caral (Adkins and Benfer 2009) and been suggested in Chankillo (Ghezzi and Ruggles 2015) and other later cultures of coastal Peru, such as the Chimu and Moche (Golte 2009; Makowski 2010; Urton 1982). However, a detailed analysis focusing on cultural astronomy has seldom been conducted in these areas, and relevant evidence has only been outlined for the sites of Huánuco Pampa (Pino Matos 2004), Intimachay in Machu Picchu (Ziólkowski et al. 2015), and the southern Andes (see Moyano 2016 and references therein), 4,000 years later than the earliest structures in the Supe Valley.

The Caral people's suggested interest in the moon was probably related to its southernmost moonrise, notably at full moon, which was perhaps realized as an important land and time marker. This coincided with the start of the winter months for this area (roughly from May to October; Figure 8). Interestingly, this fits the end of one of the present-day seasons to collect anchoveta (*E. ringens*), one of the main sea products recovered at the Supe Valley archaeological sites (Burger and Rosenswig 2012; Shady 2006a). It would also coincide with the time to sow cotton, one of the most important products of the middle section of the valley where the most prominent buildings are located. In addition, the heliacal rising of Sirius at the time of use of these structures would happen in June, perhaps coincidentally with the lunar event. Finally, the heliacal rising of the Southern Cross at those times would happen in September, a few weeks earlier than the start of the flooding in mid-spring and marking the beginning of the pumpkin cycle.

<Insert Figure 8 about here>

Our analysis also suggests that a number of structures could target the solstitial positions of the Sun. This could be of importance as the full Moon extremes happen around the solstices. Indeed, several structures in Aspero, Lurihuasi, Miraya, and Caral seem to present the possibility of a double alignment toward the moon and the sun. Even though the results of the statistical analyses indicate the predominantly eastern directionality of orientations, the possibility should not be discarded that some of them were (also) functional to the west, suggesting that both events—the major lunistice and the

solstice—were observed during the same period of the year. A similar case could be indicated in Mesoamerica for some sites in the northeastern coast of Yucatan (Sanchez Nava et al. 2016)

In summary, these orientations might indicate the prominence of particular moments that could be related to how the Caral society envisioned their land and time in a coherent manner, indicating the interrelationship of the moon, the river, and economic activities. As indicated by Tilley (1984:122), "places stand out as vested with meaning and significance." Monuments such as megaliths "deny time" with their permanence and stability, attributes that could be common to most monumental architecture, including pyramidal buildings. When we find, moreover, a coherent set of orientations like those in the Supe Valley, we can conclude that they deny time by incorporating its circular flow: they embed a cyclic perception of time. Such is provided by the timely recurrence of astronomical events, such as the rising and setting of the sun or the moon at certain moments of the year.

It would be simplistic to suggest that these recurrences account for the existence of astronomical observatories (in the Western sense of sites to perform scientific investigations) in each monumental pyramidal building in the Supe Valley. Rather, we argue that these relationships to the river and the sky point toward the ritual importance of having the monuments built at the right places, with the correct orientation with respect to the river and in concordance with the movements and rhythms of nature—particularly the sky, a model that is more complex and culturally meaningful. The monuments' placement seems to indicate the importance of the moon for this society for timekeeping, possibly not only to mark economic periods but also to identify the moments to begin the particular rituals associated with the start or end of some economic activities in a way that might have reinforced social cohesion (Ziedler 1998).

In such a way, the monumental architecture of the Supe Valley as a product of social action and a constraint on future action (Tilley 1984:134) includes this temporality. It might have started with the impressive labor of building those large structures, but could have been maintained over time with their continuous remodeling to enable the reenactment of rituals at particular times of the year—in this case perhaps dictated by the visibility of the full moon closest to the JS and in line with one of the main axes of those buildings. The sky is an integral part of the archaeological landscape (Sprajc 2018) and therefore has a specific meaning (Ashmore 2008), providing clues to complement our understanding of landscape organization in the past (Knapp and Ashmore 1999). The interaction of the material remains with the landscape of the Supe Valley, and in particular with how the river and the sky can be perceived from them, provides hints to its social meaning (Criado-Boado 2014) and the interwoven dimension of space and time in social systems (Iwaniszeski 2003).

Supe society had access to the resources of one of the world's most productive seas, and the low section of the Supe Valley afforded favorable conditions for agriculture. These conditions, facilitated by the annual river floods, drove a flourishing economy that indeed became petrified in the orientation of their public buildings, revealing how they looked to the sky in a search for signals to coordinate these social activities.

*Acknowledgments.* We thank Prof. Clive Ruggles, Dr. Ivan Ghezzi, and the several referees for valuable comments and suggestions that helped improve the quality of this article. This work was partially supported by the projects P/310793 “Arqueoastronomía” of the IAC, AYA2015-66787P “Orientatio ad Sidera IV” of the Spanish MINECO, and the AECID grant “Astronomía en la Cultura del Valle del Supe, Perú” from the Spanish MAEC. This research was carried out within the framework of the Archaeological Research Program, “Caral Civilization in the Valleys of Supe and Huaura during the Initial Formative and Early Formative” of the Caral Archaeological Zone, Executing Unit No. 003 of the Ministry of Culture of Peru, which since 2014 has been studying mainly the orientation of the buildings of the Caral archaeological site. We acknowledge support of the publication fee by the CSIC Open Access Publication Support Initiative through its Unit of Information Resources for Research (URICI).

*Data Availability Statement.* All data are available in the main text.

**Notes**

1. The astronomical declination, usually noted as δ, is a celestial coordinate that allows possible astronomical referents of alignments to be determined. It has the advantage that a single magnitude allows direct comparison of measurements at different sites. It should be emphasized that an azimuth corrected for the altitude of the horizon does not have the same effect, because it is still affected by latitude changes, whereas δ is not. To obtain δ we need the measurement of the azimuth (A), the altitude of the horizon (h), and the latitude of the site (φ). The relation between the three is given by the formula:

$$\sin \delta = \sin h \sin \varphi + \mathrm{con}\, h \cos \varphi \cos A$$

The declination of the sun at the equinoxes is 0°, defining the equinox as the moment when the sun crosses the celestial equator. The declination of the sun in the solstices during the change of Era was around ± 23.7° (whereas today it is ± 23.4°). It should be noted that the moon varies its position between a series of limits located a few degrees north and south of the sun's own values (c. ± 29°). Finally, for a given moment, the stars have a constant astronomical declination, although such magnitude is subject to change due to the secular movement of the Earth axis called the precession of the equinoxes. All these changes have been considered in our analysis.

2. The orbit of the moon has an angle of inclination, *i*, of nearly 5° 9' with respect to that of the ecliptic (Williams and Dickey 2003). This means that the orbital plane of the moon and that of the Earth intersect in a line, called the line of nodes. Tidal forces provoke a retrograde motion of that line of nodes, with a period of 18.6 years (Williams and Dickey 2003). This movement has an effect on how moonrise or set is seen on any spot on the surface of the Earth, reaching two extremes in that period, with declinations ± $|\varepsilon + i|$, where $\varepsilon$ is the obliquity of the ecliptic. Alexander Thom coined the term "major lunar standstill" for the most extreme declinations (a.k.a. major lunistice), which would happen for values close to ± 29º, whereas the minor lunar standstill would happen for values close to ± 18º (González-García 2015).

Figures

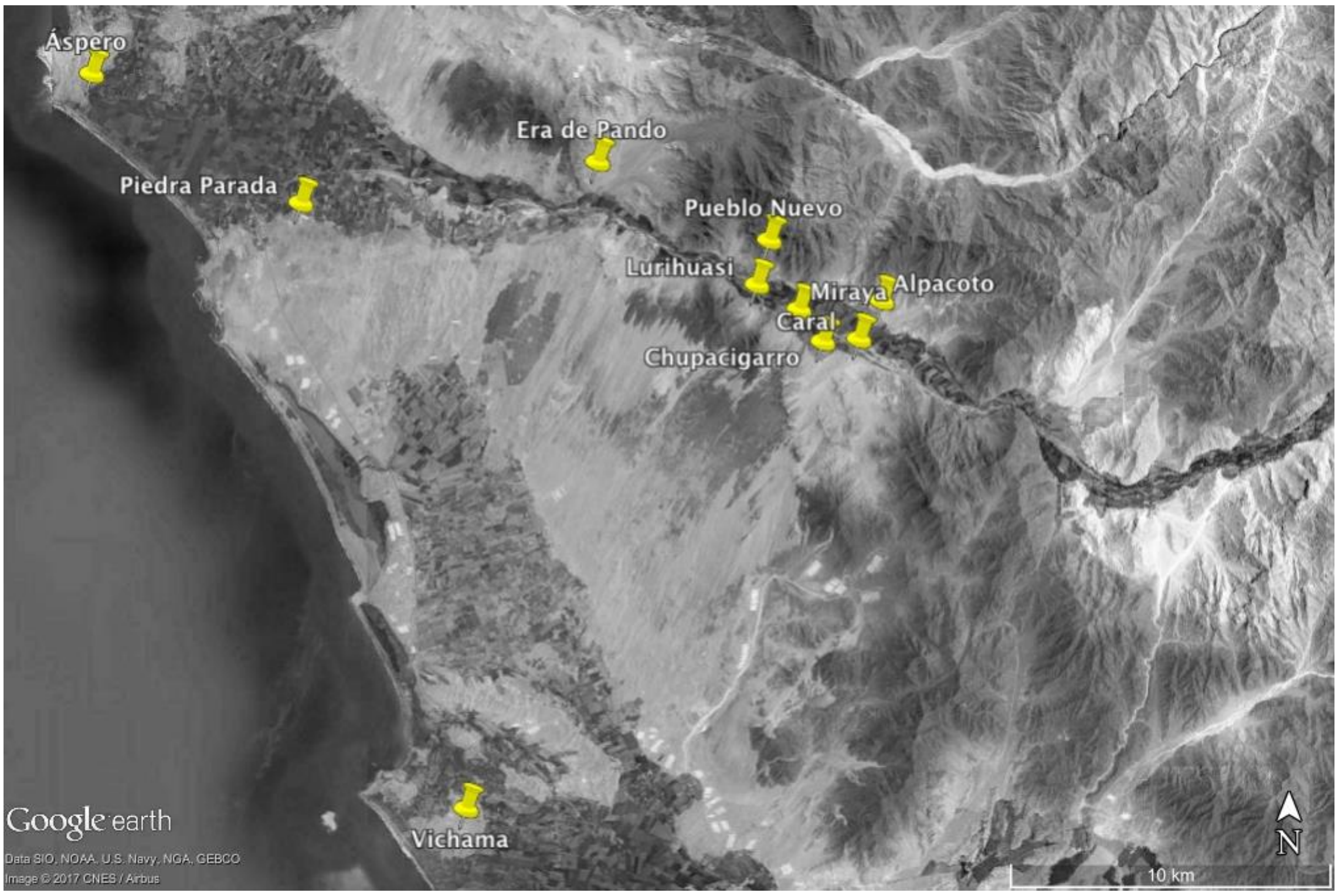

Figure 1. Location of the Supe Valley in Peru's north-central area and the sites analyzed in this article within or close to it. Note the direction of the riverbed at those sites. Two sites are outside the Supe Valley: Piedra Parada next to the seashore, and Vichama, to the south in the Huaura Valley.

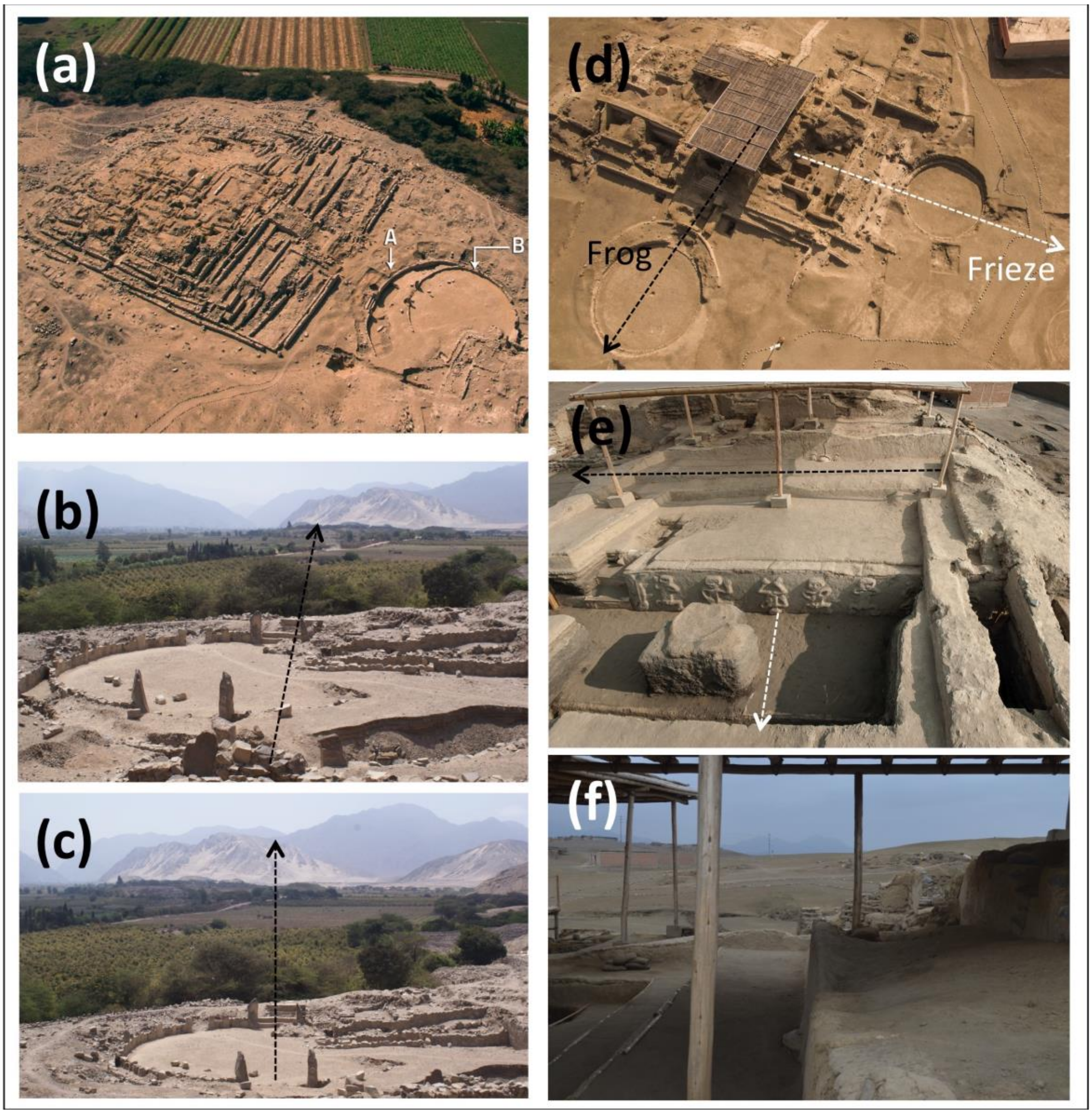

Figure 2. Examples of sunken circular plazas. (a) Aerial view of building A in Miraya. Note the change in size and orientation between the oldest (and largest, A) and newest (B) *plazas*. Whereas building A is facing toward Caral in the background (notice the top of buildings C and E of Caral in inset (b)) and the area where the southernmost moonrise would happen, the B *plaza* (c) is facing toward a different area, allegedly related to the rise of the Southern Cross. (d) Building A1 in Vichama has two sunken circular plazas of different epochs. This building is famous for the appearance of the dancing figures in a frieze and the figure of a frog (e). The oldest plaza—white arrow in (d)—presents an orientation in coincidence with the direction of the frieze and the frog wall. The newest plaza, built at a later period, is not exactly perpendicular to the oldest one (f). The orientation of both plazas towards southeast would coincide with the rising of Sirius, the brightest star in the sky, for the two periods of construction on site during the second millennium BC.

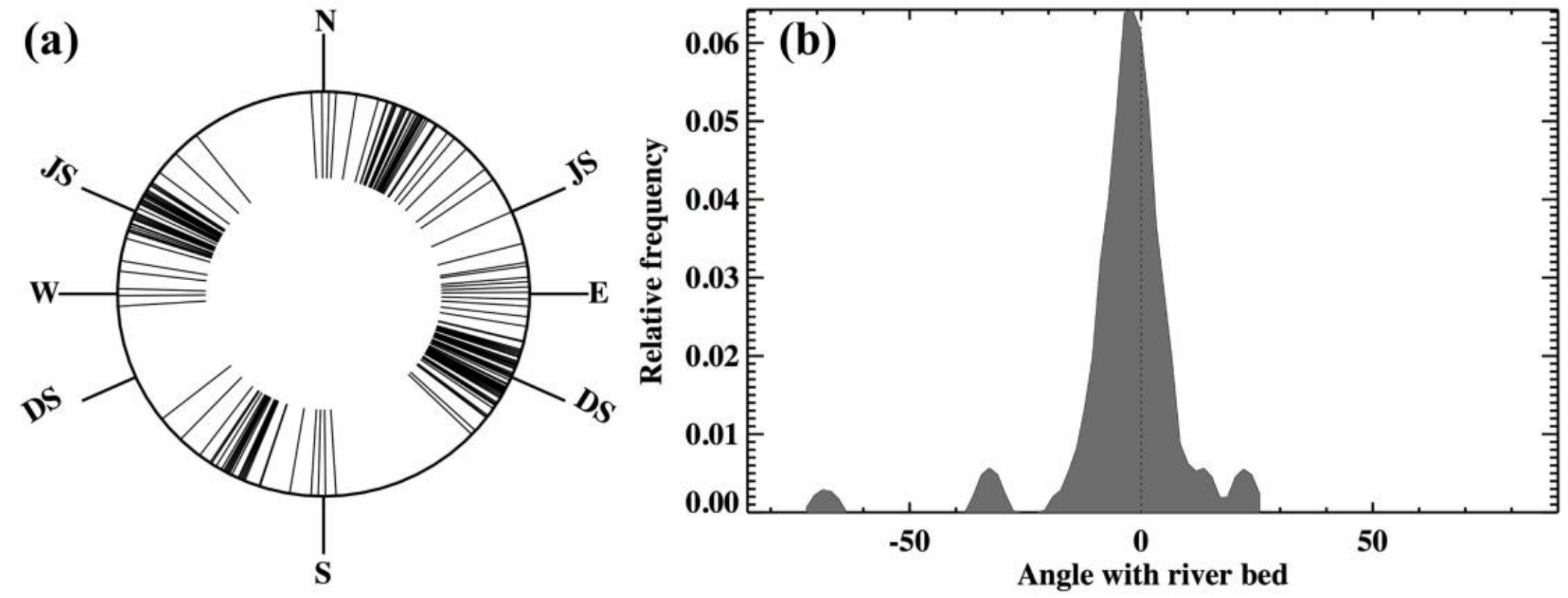

Figure 3. (a) Orientation diagram for all structures measured and included in Table 1. Each measurement is indicated as a linear stroke inside the circle. The solid lines outside the circle indicate the cardinal points and the extreme positions of sunrise in the solstices (JS stands for June solstice and DS for December solstice) at the latitude of the Supe Valley. (b) Frequency histogram showing the angle of the structures with respect to the riverbed closer to them. The largest concentration indicates that most of the buildings follow the river flow quite closely.

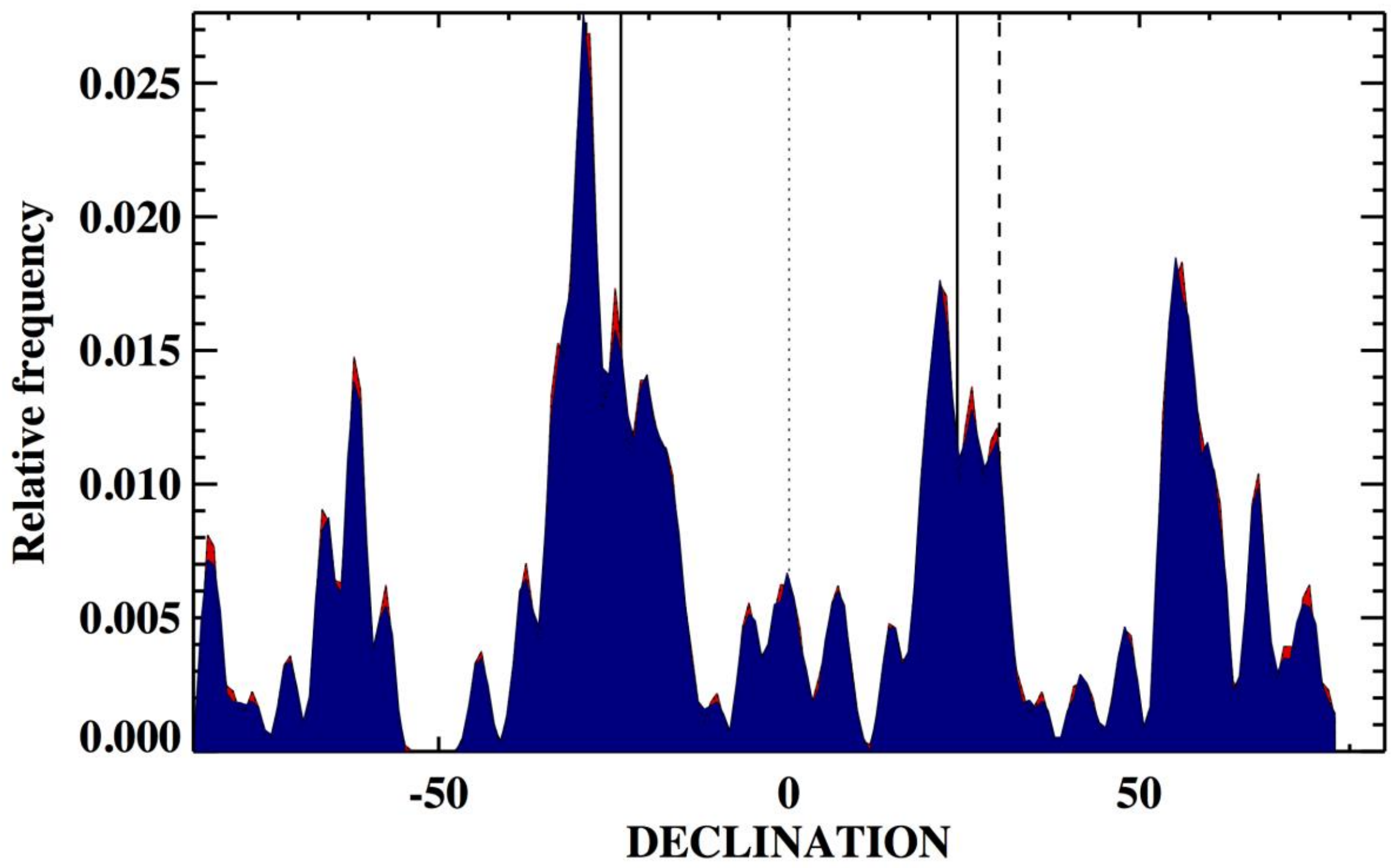

Figure 4. Declination histogram built to compare the effect of using different kernels. The red area uses a Gaussian kernel, whereas the dark blue employs an Epanechnikov. Vertical dashed lines stand for the extreme positions of the moon, vertical solid lines indicate the solstices, and the dotted vertical line indicates the astronomical equinox.

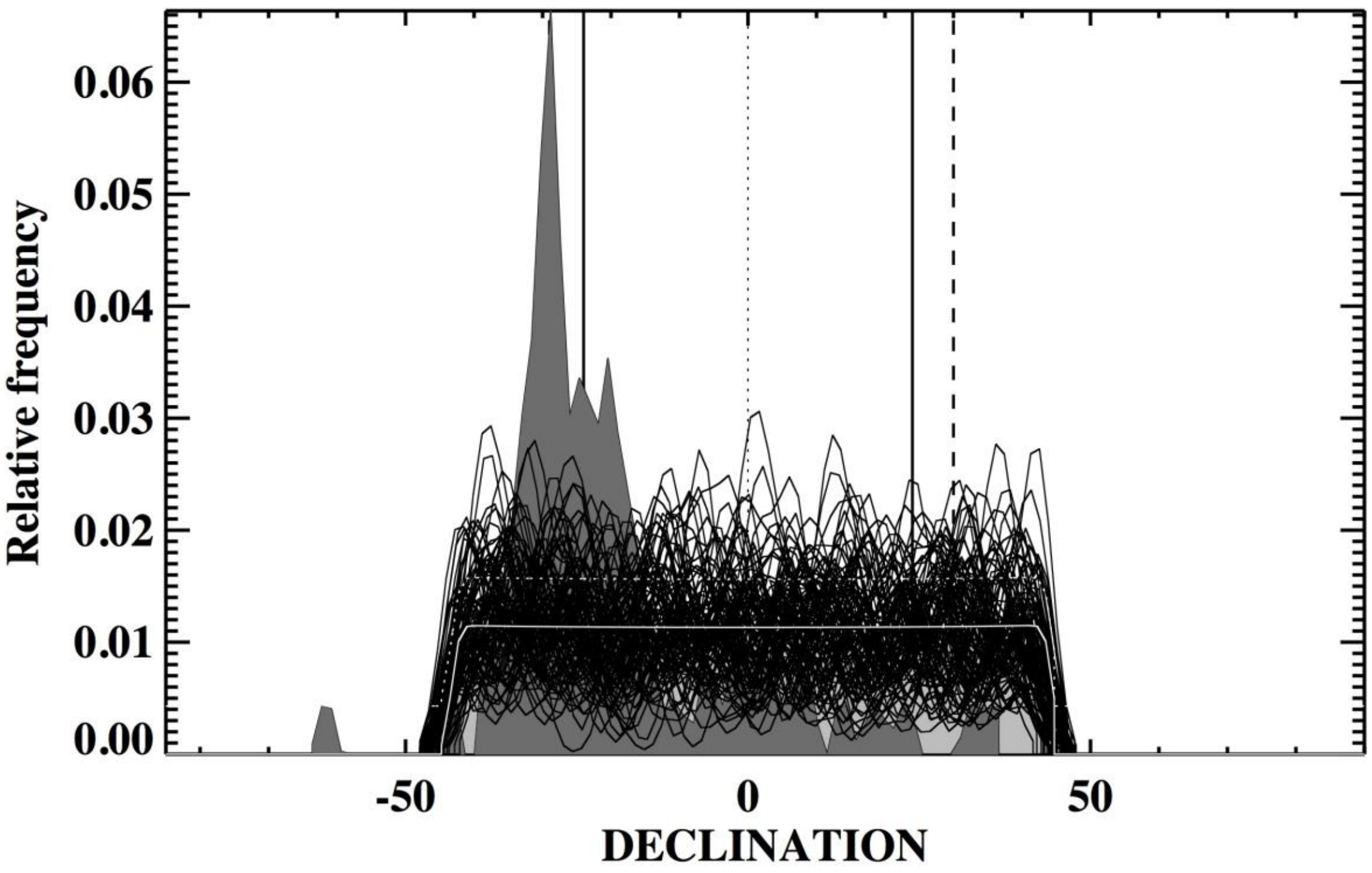

Figure 5. The relative frequency diagram of the eastern data (dark gray) is compared with 100 distributions randomly chosen from a pool of 80,000 uniformly distributed orientations within this eastern sector (i.e., from 45º to 135º); each of the 100 random distributions are given by the black solid lines. The observed distribution is then compared with that uniform distribution (solid white line). The standard deviation of the 100 random distributions is computed (upper dotted white line), and any peak three times more prominent than this standard deviation is considered significant. Vertical lines are as in Figure 4.

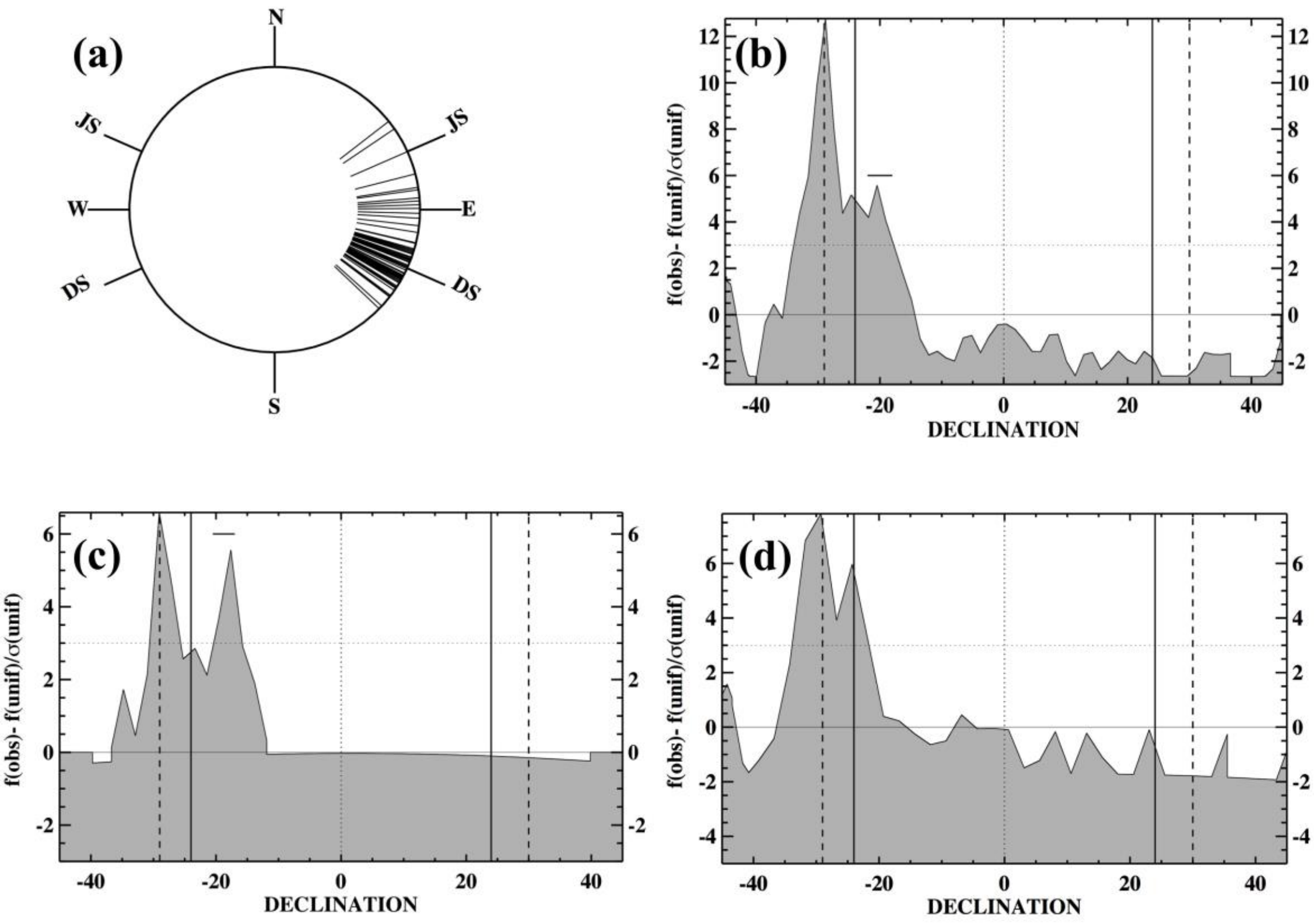

Figure 6. Results considering the eastern horizon. (a) Orientation diagram of measurements in Table 1 toward the eastern sector of the horizon. Note the concentrations around the December solstice (DS) sunrise. (b) Histogram of declination of all buildings toward the eastern sector. Vertical lines are as in Figure 4. The horizontal dotted line at a value equaling 3 indicates the level of 3-sigma. Note the concentration toward $\delta = -29°$, corresponding to the southernmost rising of the moon. Secondary maxima appear at values $\delta = -24°$ and $\delta = -20°$, possibly related to the DS sunrise and perhaps the minor lunistice or the rise of Sirius, the brightest star in the night sky. The horizontal solid stroke indicates the change in declination of Sirius for the period between 3000 and 1500 BC. (c) Declination histogram of the eastern orientations of the sunken circular plazas. Note the prominence of the orientations toward $\delta = -29$ and $\delta = -18°$. In addition to a possible third concentration toward the DS sunrise ($\delta = -24°$), there is a fourth, less significant concentration well below our significance level toward $\delta = -34°$, tentatively related to the rising of the Southern Cross. The horizontal solid stroke indicates the change in declination of Sirius for the period between 2400 and 1500 BC. (d) Same but only including buildings not associated with circular plazas; notice how the peak at c. −18/20° disappears while the solstice signal is reinforced.

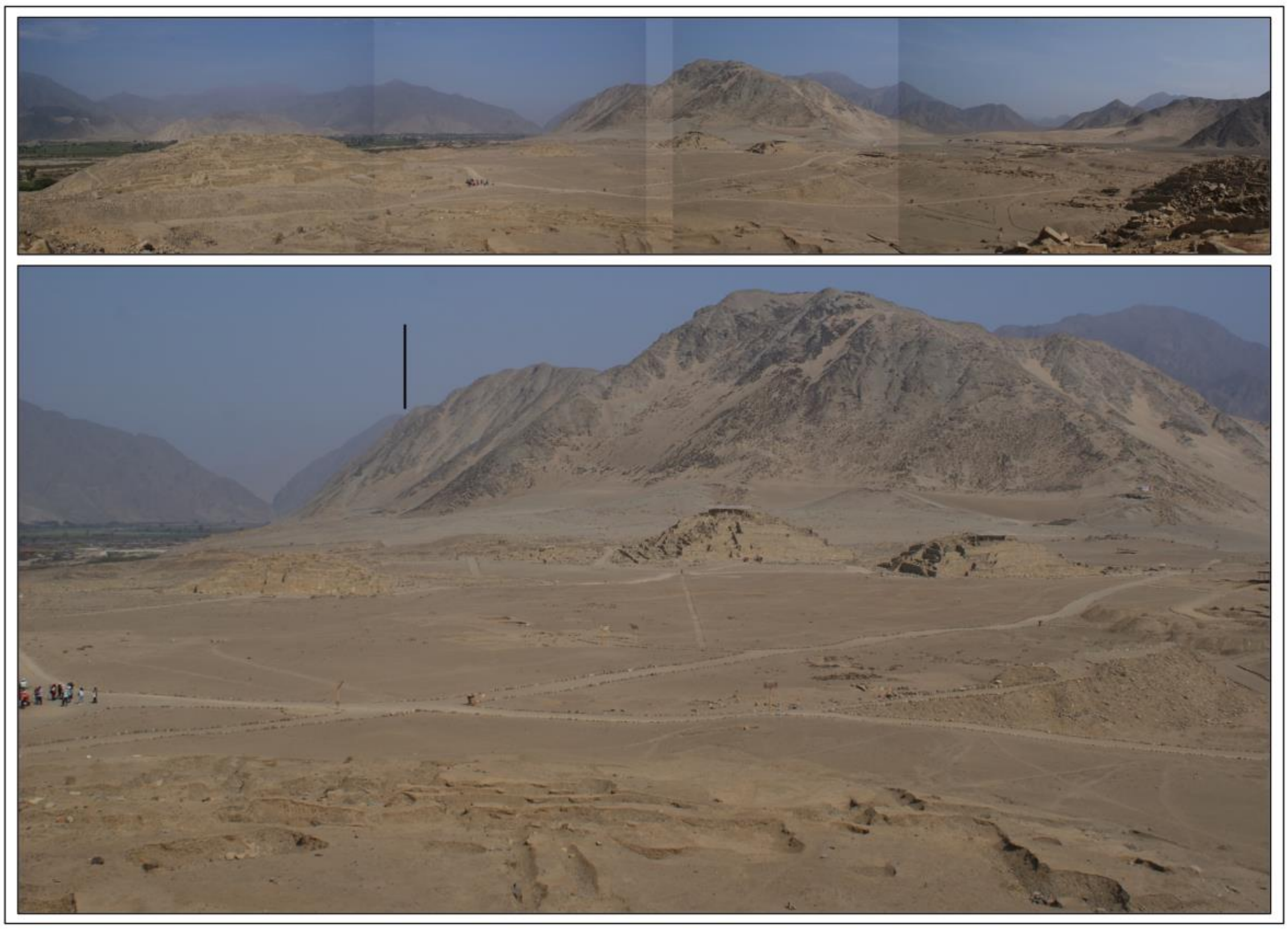

Figure 7. Panoramic view of Caral as seen from the top of building C1 toward the eastern part of the horizon (top). This pyramidal building houses several interesting orientations, very similar to other buildings in Caral, such as building E (seen on the left of the panorama). The bottom is a close-up view, with a vertical line indicating the orientation of this building toward the DS sunrise.

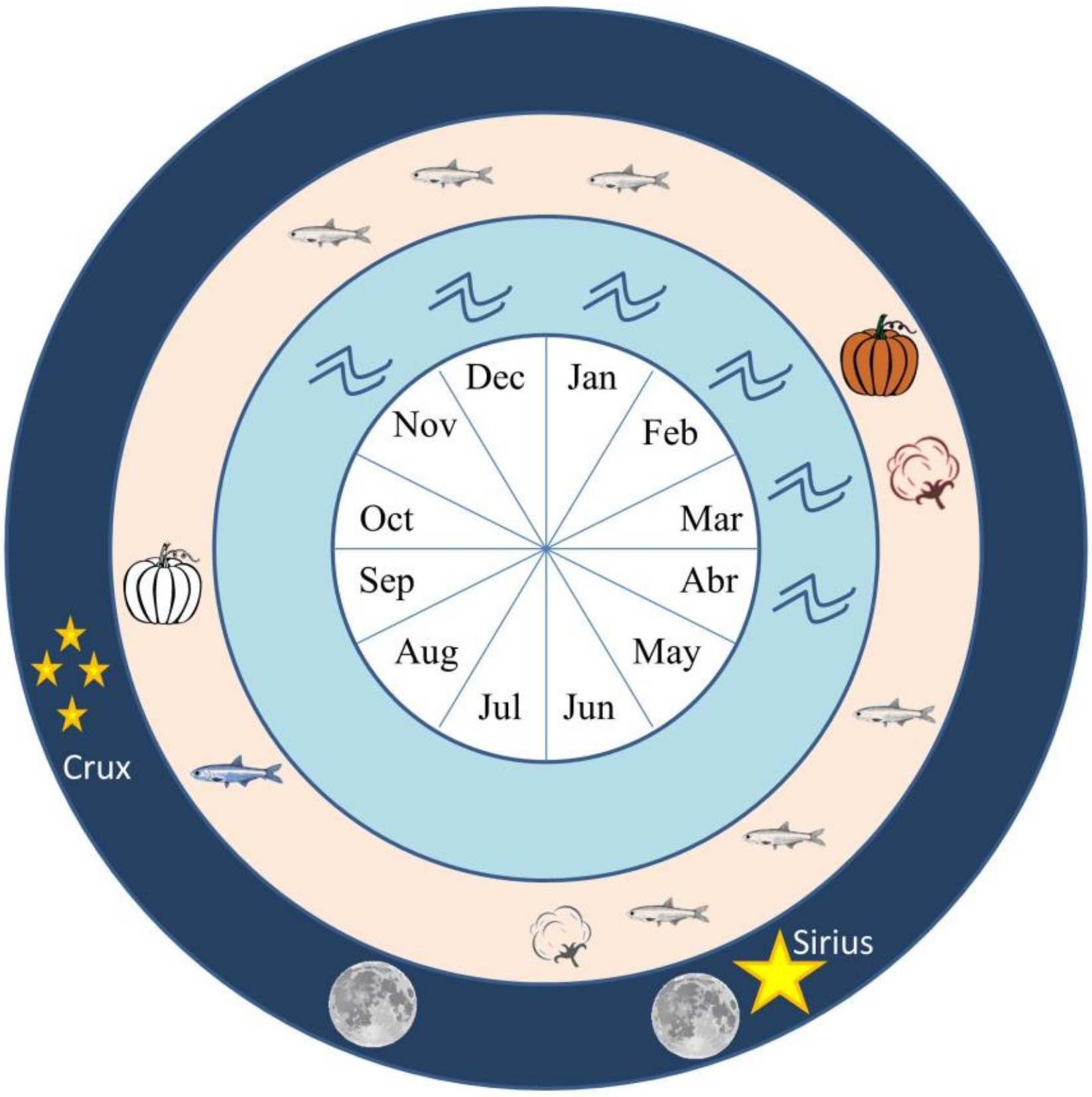

Figure 8. Chronogram of different events in the Supe Valley. The inner circle indicates the months for temporal reference. The light blue circle indicates the period when the River Supe increases its flow due to the rains in the high Andes (waves). The light brown circle indicates some productive cycles in this part of the Andes. Black-and-white symbols in this circle indicate the period of sowing, whereas color symbols stand for harvest time. Only cotton and pumpkin are considered here. The black-and-white fish symbol indicates the two seasons during which fishermen in Peru currently collect anchovy; the color symbol indicates the current spawning period. Finally, the outer dark blue circle indicates the relevant astronomical phenomena indicated by the orientation of the measured structures. The two moons indicate the period of full moon visibility before and after the June solstice. These are the moons that would appear in the southeast section of the horizon. The stars indicate the heliacal rising of Sirius (big star) happening at the beginning of June and the heliacal rising of the Southern Cross (group of four stars) at the beginning of September for the time of Caral.

Table 1. Orientation of 55 Structures at 10 Sites in the Supe Valley, Peru.

| Site | Monument | φ (º/’) | λ (º/’) | a (º) | h (º) | δ (º) | Commentaries |
|---|---|---|---|---|---|---|---|
| Aspero | Huaca ídolos | -10/48 | 77/44 | 85¼<br>81 | 2¼<br>1 | 4.3<br>8.7 | <br>Sunken plaza |
| | Huaca alta | | | 55½ | 4 | 32.9 | |
| | Plaza hundida | | | 39½<br>119½ | 3½<br>0¾ | 48.2<br>-29.0 | Stairs<br>perpendicular |
| | Espondilus | | | 75½<br>82 | 2½<br>2 | 13.8<br>7.5 | Stairs 1<br>Stairs 2 |
| | Huaca sacrificios | | | 66 | 3¼ | 22.9 | |
| Era de Pando | C1 | -10/50 | 77/35 | 179½<br>89½<br>269½<br>359½ | 3¾<br>8¼<br>14¾<br>5½ | -82.7<br>-1.04<br>-3.2<br>73.8 | |
| | Sunken plaza C1 | | | 179½ | 3¾ | -82.7 | Notch 86¾; h = 7; δ = 1.9 |
| | B | | | 93½<br>3½<br>183½ | 8½<br>4½<br>4 | -5.0<br>74.4<br>-82.2 | |
| | A1 | | | 9¼<br>99¼<br>279½<br>189½ | 6½<br>7¾<br>---<br>6½ | 70.5<br>-10.5<br>6.2<br>-79.6 | <br><br>Assuming h~15<br> |
| Pueblo Nuevo | F | -10/51 | 77/32 | 208¼ | 5¾ | -61.6 | |
| | Sunken plaza F | | | 208¼<br>118¼ | 5¾<br>5 | -61.6<br>-28.6 | |
| | H | | | 123¼<br>117½<br><br>127½ | 3½<br>5½<br><br>5 | -33.3<br>-28.0<br><br>-37.7 | Stairs 1<br><br><br>Secondary stairs |
| | Sunken plaza H | | | 118¼ | 4¾ | -28.6 | |
| Piedra Parada | A1 | -10/51 | 77/40 | 20¼<br>110¼<br>290¼ | 3<br>4<br>-0¼ | 65.7<br>-20.6<br>20.05 | |
| | Sunken plaza A1 | | | 18¼<br>108¼<br>198¼<br>288¼ | 2¼<br>5¾<br>12½<br>-0¼ | 67.8<br>-18.9<br>-72.06<br>18.09 | |
| | B1 | | | 91½<br>181½<br>271½<br>1½ | 2¼<br>4<br>1¼<br>2½ | -1.8<br>-82.8<br>1.3<br>76.8 | |
| Lurihuasi | A3 | -10/52 | 77/33 | 321¾<br>51¾<br>231¾ | 4¼<br>5½<br>9½ | 49.0<br>36.0<br>-39.1 | |
| | D1 | | | 356½<br>86½<br>266½<br>176½<br>88 | 5¼<br>12½<br>14½<br>24<br>12½ | 73.7<br>1.03<br>-6.03<br>-76.5<br>-0.4 | <br><br><br><br>Exterior wall |
| | E2 | | | 26½<br>116½<br>206½<br>296½ | 11¾<br>8¾<br>12½<br>16 | 55.3<br>-27.5<br>-64.01<br>21.68 | |
| | F1 | | | 37<br>127<br>217<br>307 | 12<br>6¾<br>21<br>18½ | 46.8<br>-37.5<br>53.1<br>30.06 | |
| | H1 | | | 109½<br>111<br>20¼<br>299½<br>301<br>119½<br>103½ | 7½<br>6¼<br>12½<br>22½<br>22½<br>9¼<br>10½ | -20.5<br>-21.7<br>59.2<br>22.01<br>23.3<br>-30.5<br>-15.05 | <br>Twisted axis<br><br><br><br>Southern building<br>Northern building |
| Allpacoto | C1 | -10/52 | 77/30 | 211½<br>207½ | 4<br>3½ | -58.0<br>-61.7 | Stairs<br>Perpendicular |
| | | | | 117¾<br>297¾ | 7¾<br>3¼ | -28.6<br>26.5 | Pyramid |
| | B2 | | | 288¾ | -0½ | 18.6 | |
| Miraya | A1 | -10/52 | 77/32 | 123<br>33<br>303<br>213 | 4¾<br>4¼<br>8¼<br>12¼ | -33.2<br>53.9<br>30.2<br>-57.7 | East façade stairs |
| | | | | 28¾<br>118¾<br>208¾<br>298¾ | 7<br>4½<br>11½<br>10 | 56.3<br>-29.03<br>-61.8<br>25.7 | North façade stairs |
| | Sunken Plaza A1 | | | 116½<br>124½ | 4½<br>4¾ | -26.8<br>-34.7 | Large sunken plaza<br>Small sunken plaza |
| | C1 | | | 107½<br>28½ | 3¾<br>8 | -17.8<br>56.0 | Old structure<br>Stairs |
| | C3 | | | 118<br>208<br>298<br>28 | 4½<br>12<br>10<br>4½ | -28.3<br>-62.5<br>24.9<br>58.3 | |
| | C4 | | | 33¼<br>122½<br>302½ | 4¼<br>4½<br>8½ | 53.7<br>-32.7<br>29.6 | W Building |
| | | | | 33½<br>123½<br>303½<br>213½ | 4½<br>5<br>7½<br>15¾ | 53.4<br>-33.8<br>30.9<br>-57.08 | E Building |
| | C5 | | | 30½<br>120½<br>300½<br><br>123½ | 4½<br>4½<br>9<br><br>4½ | 56.05<br>-30.7<br>27.6<br><br>-33.7 | <br><br><br><br>Stairs |
| | Sunken plaza B1 | | | 33¼<br>123¼ | 3¾<br>5 | 53.9<br>-33.5 | |
| Caral | C1 | -10/53 | 77/31 | 114<br>114 | 5¾<br>7 | -24.6<br>-24.8 | Up<br>Bottom |
| | C2 | | | 209<br>119<br>29 | 8½<br>8 | -61.3<br>-29.8 | <br>Perpendicular |
| | B10<br>La Cantera | | | 27½<br>117½<br>30½<br>120½<br>119¼<br>299¼<br>29¾<br>120<br>300 | 6½<br>10<br>5½<br>9¾<br>9¾<br>1¼<br>5<br>10½<br>1 | 57.6<br>-28.6<br>55.6<br>-31.5<br>-30.3<br>28.5<br>56.7<br>-31.1<br>29.2 | Stairs 1st building<br>Perpendicular<br>Stairs 2nd building<br>Perpendicular<br>Stairs final period<br>+180º<br>Late period stairs<br>To the top of mountain<br>+180º |
| | Circular structure with oven | | | 121 | 10½ | -32.1 | |
| | P | | | 134<br>314<br>44<br>224 | 8¾<br>B<br>6¾ | -44.7<br><br>42.8 | Circular altar<br><br><br>Perp. 131 / 7 ; $\delta$ -41.5 |
| | Amphitheater | | | 29¾<br>119¼<br>209¾<br>299¾ | 7½<br>12¼<br>8<br>2 | 55.2<br>-30.6<br>-60.5<br>28.8 | Sunken plaza |
| | Next to amphitheater | | | 28½<br>118½<br>116¾<br>296¾ | 7<br>12<br>12<br>2¼ | 56.5<br>-29.8<br>-28.1<br>25.8 | <br><br>Outer wall |
| | Fire altar | | | 297½ | 2½ | 26.5 | |
| | P. Huaca | | | 23¾<br>113¾<br>293¾ | 5½<br>11½<br>2¼ | 61.3<br>-25.2<br>22.9 | Stairs<br>Wall |
| | Huaca | | | 207¼<br>132½ | 6¼<br>15¼ | -62.6<br>-43.6 | Mountaintop P. Huaca<br>Mountaintop P. Galeria |
| | P Galeria | | | 290½ | 3½ | 19.4 | Stairs |
| | | | | 110½ | 8¼ | -21.5 | +180º, h hasta 12½ |
| | | | | 292 | 4¼ | 20.7 | earlier phase |
| | | | | 25½ | 6½ | 59.3 | gallery |
| | | | | 203¼ | 4¾ | -66.1 | closed stairs |
| | F8 | | | 114 | 6¾ | -24.7 | Squared plaza |
| | | | | 292½ | 2½ | 21.6 | |
| | | | | 204½ | pir. | | |
| | | | | 23½ | 6½ | 60.9 | |
| | G1 P Menor | | | 293½ | 2¼ | 22.6 | |
| | | | | 113½ | 6 | -24.1 | +180º |
| | | | | 27¼ | 6¾ | 57.7 | Previous phase |
| | | | | 297 | 1½ | 26.2 | Room |
| | | | | 117½ | 6½ | -28.1 | |
| | P Principal | | | 202½ | 5¼ | -67.06 | Sunken plaza |
| | | | | 22½ | 11 | 58.8 | Perp. toward pyramid |
| | | | | | 17 | 54.4 | middle plaza |
| | | | | | 33½ | 40.7 | stairs |
| | | | | 111¾ | 5 | -22.2 | Perpend. |
| | | | | 119 | 7½ | -29.8 | Pyramid base line |
| | | | | 291¾ | 2 | 21.0 | Top part stairs |
| | | | | 204¼ | 4½ | -65.1 | Room |
| | | | | 295¼ | 1½ | 24.9 | |
| | | | | 114½ | 5¼ | -25.0 | |
| | | | | 25 | 5 | 60.6 | To the river |
| | E2 | | | 293 | 2 | 22.2 | |
| | | | | 23 | 6½ | 61.3 | |
| | | | | 113 | 5¾ | -23.6 | |
| | | | | 203 | 6 | -66.7 | |
| | 'Observatory' | | | 111½ | 6 | -22.2 | |
| | | | | 115¼ | 6¾ | -26.0 | |
| Chupacigarro | A | -10/53 | 77/31 | 18 | 3¾ | 67.0 | |
| | | | | 108 | 4½ | -18.5 | |
| | | | | 288 | 13½ | 14.6 | |
| | | | | 198 | 4 | -70.8 | |
| | Sunken plaza A1 | | | 23¾ | 3¼ | 62.6 | |
| | | | | 113¾ | 6 | -24.4 | |
| | | | | 293¾ | 11¾ | 20.4 | |
| | | | | 203¾ | 7½ | -66.3 | |
| Vichama | A1 | -11/01 | 77/38 | 19¾ | 1¾ | 66.8 | Old sunken plaza stairs |
| | | | | 109½ | 2 | -19.5 | Perp. |
| | | | | 20¾ | 1¾ | 65.9 | North façade |
| | | | | 110¾ | 1¾ | -20.6 | |
| | | | | 109¾ | 2 | -19.7 | Perp. |
| | | | | 289½ | 0 | 19.2 | Frize |
| | | | | 106 | 2 | -16.03 | Toad |
| | | | | 286 | 0 | 15.8 | |
| | | | | 107 | 1¾ | -17.0 | New sunken plaza stairs |
| | | | | | 3½ | -17.3 | Base |
| | K | | | 96½ | 1 | -6.5 | |
| | | | | 276½ | 0 | 6.6 | |
| | E2 | | | 15½ | 0 | 71.4 | Trapezoidal building |
| | | | | 20½ | 0 | 67.1 | |
| | Sunken plaza C1 | | | 127¾ | 0¾ | -37.01 | |
| | D1 | | | 106½ | 1¾ | -16.6 | |
| | Sunken plaza D | | | 103¾ | 2¼ | -13.9 | |

*Notes:* For each monument, the location, identification of the structure, latitude and longitude of the site (φ and λ), its azimuth (a), the angular height of the horizon (h) in that direction, and the corresponding declination (δ) are shown. The last column contains some additional comments or data for alternative orientations (in º).